\documentclass[11pt,reqno]{amsart}
\usepackage[margin=1.2in]{geometry}
\usepackage[usenames,dvipsnames]{xcolor}
\definecolor{linkColor}{HTML}{f10000}
\usepackage[colorlinks,allcolors=linkColor]{hyperref} 
\usepackage{amsmath, amssymb, amsthm,mathrsfs}
\usepackage{booktabs}
\usepackage{array}
\usepackage[pdf]{pstricks}
\usepackage{pst-plot,pst-node,pst-func,pst-grad}
\usepackage{multido}
\usepackage[nomessages]{fp}
\usepackage{cleveref}
\usepackage{setspace}
\usepackage{soul}
\usepackage{cite}

\newtheoremstyle{indented}{3pt}{3pt}{\addtolength{\leftskip}{1.5em}
        \itshape}{}{\bfseries}{.}{.5em}{}

\theoremstyle{indented}
\newtheorem{remark}{Remark}
\AtBeginDocument{
  \addtocontents{toc}{\small}
  \addtocontents{lof}{\small}
}
\newcommand{\dt}{\frac{d}{dt}}
\newcommand{\dx}{\frac{d}{dx}}

\newcommand{\grad}{\nabla}
\newcommand{\D}{\mathscr{D}}
\newcommand{\vn}{\vec{\mathbf{n}}}
\newcommand{\vt}{\vec{\mathbf{t}}}
\newcommand{\dotn}{\cdot\vn}
\newcommand{\dott}{\cdot\vt}
\newcommand{\normD}[1]{\frac{\partial#1}{\partial\vn}}
\newcommand{\cint}[1]{\oint_{\partial\D}\left[#1\right]\,ds}
\newcommand{\sint}[1]{\int_{\mathscr{S}}{\left[#1\right]\,dx}}
\newcommand{\bint}[1]{\int_{\mathscr{B}}{\left[#1\right]\,dx}}
\newcommand{\wint}[1]{\int_{-\infty}^\infty{\left[#1\right]\,dx}}
\newcommand{\fint}[1]{\int_{-\infty}^\infty{e^{-ikx}\left[#1\right]\,dx}}

    \title{A Weak Formulation of Water Waves In Surface Variables}
    \author{KL Oliveras}
    \date{\today}
    \address{Mathematics Department, Seattle University, Seattle, WA 98122}
    \email{oliveras@seattleu.edu}
    \urladdr{fac-staff.seattleu.edu/oliveras/web} % Delete if not wanted.
\begin{document} 
\begin{abstract}
    In this short note, we derive a system of two nonlocal equations for the water-wave problem following the work of \cite{ablowitz2006new}.  Specifically, we consider a fluid with a one-dimensional free surface for an irrotational fluid both with, and without, surface tension.  We show that these equations can be useful for deriving direct maps between boundary data at the various interfaces and consider various asymptotic regimes.  Finally, for periodic traveling wave solutions, we derive a new single-equation that is derivative free and has the potential to simplify the process of finding asymyptotic expansions of the solutions as well as reduce numerical errors when solving computationally.  
\end{abstract}
\maketitle
% \begin{spacing}{0.0}
%     \small
%     \tableofcontents
% \end{spacing}

%%%%%%%%%%%%%%%%%%%%%%%%%%%%%%%%%%%%%%%%%%%%%%%%%%%%%%%%%%%%%%%%%%%%%%%%
\section{Introduction}
Various reformulations of the water-wave problem for a one-dimensional surface have been proposed.  Examples include conformal variables, boundary-integral formulations, the Zakharov-Craig-Sulem formulation \cite{zakharov1968stability,craig1993numerical}, the Ablowitz-Fokas-Musslamini formulation \cite{ablowitz2006new}.  Each formulation presents some advantages and disadvantages as discussed in \cite{wilkening2015comparison}.  

Here, we introduce yet another reformulation heavily based on the previous work of \cite{ablowitz2006new}.  While this formulation may suffer some of the same disadvantages as described in \cite{wilkening2015comparison}, the hope is that the formulation is generalized in a manner that would compensate for these numerical challenges.  One specific advantage is that for traveling-wave solutions, the formulation yields a derivative-free formulation which does appear to be more numerically stable than other formulations.  Furthermore, the formulation does allow for direct maps between the free-surface and the pressure complementing the previous work of \cite{bonneton2017recovering}.

%%%%%%%%%%%%%%%%%%%%%%%%%%%%%%%%%%%%%%%%%%%%%%%%%%%%%%%%%%%%%%%%%%%%%%%%
\section{Derivation}

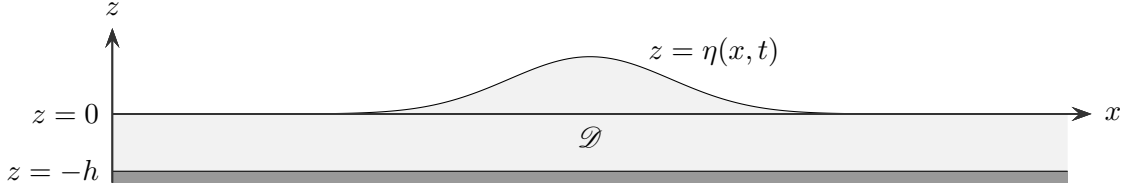
\begin{figure}[htp]
        \centering
        \psset{xunit=.5in,yunit=.3in,plotpoints=200,plotstyle=line}
        \begin{pspicture}(-6,-1)(6,1.75)
            % Plot the fluid domain
	        \pscustom[linestyle=none,linewidth=.2pt,fillstyle=solid,fillcolor=black!05,algebraic=true]{%
                \psplot{-5}{5}{-1} %
                \psplot{5}{-5}{2^(-x^2)} 
            }
            \psplot[algebraic=true,linewidth=.4pt]{5}{-5}{2^(-x^2)}
            % Plot the bottom
            \pscustom[linestyle=none,fillstyle=solid,fillcolor=black!40,algebraic=true]{%
                \psplot{-5}{5}{-1}%
                \psplot{5}{-5}{-1.2 } 
                } 
            \psplot[algebraic=true,linewidth=.4pt]{-5}{5}{-1}
            % Plot the axes
            \psaxes[ticks=none,labels=none,arrowscale=2,linewidth=.6pt,linecolor=black!80]{->}(-5,0)(-5,-1.2)(5.25,1.5)
            \uput[d](0,0){$\mathscr{D}$}
            \uput[r](5.25,0){$x$}
            \uput[u](-5,1.5){$z$}
            \uput[l](-5,-1){$z = -h$}
            \uput[l](-5,0){$z = 0$}
            \uput[u](1.3,.6){$z = \eta(x,t)$}
        \end{pspicture}
        \caption{Fluid Domain}\label{fig:fluidDomain}
    \end{figure}

    We being with the following formulation of the problem as illustrated in \Cref{fig:fluidDomain}:
        \begin{align}
            &\phi_{xx} + \phi_{zz} =0,  & &(x,z)\in\mathscr{D}, \label{eqn:laplace1d} \\ 
            &\phi_t + \frac{1}{2}\vert\nabla\phi\vert^2 + gz + \frac{p}{\rho} = 0,  & &(x,z)\in\mathscr{D},\label{eqn:bernoulliBulk} \\  
            &\phi_z =0, & &z = -h, \label{eqn:kinematicBottom1d}  \\
            &\eta_t + \phi_x\eta_x =\phi_z,  &&z = \eta(x,t),\label{eqn:kinematic1d}  \\
            &\phi_t + \frac{1}{2} \left(\phi_x^2 + \phi_z^2\right) + g\eta =\frac{\sigma\eta_{xx}}{\rho(1 + \eta_x^2)^{3/2}}, && z = \eta(x,t), \label{eqn:dynamic1d}
        \end{align}
    where $\phi(x,z,t)$ is the velocity potential, $\eta(x,t)$ is the free-surface deviation from a state of rest, $h$ is the undisturbed fluid depth, $g$ is the acceleration due to gravity, $p(x,z,t)$ is the fluid pressure, $\rho$ is the fluid density, and $\sigma$ is the coefficient due to surface tension. 
    
    Furthermore, we impose that both $\eta(x,t)\to 0$ and $\phi(x,z,t)\to 0$ sufficiently fast as $\vert x \vert \to \infty$. Typically, we define a velocity potential $\tilde{\phi}$ such that $\nabla\tilde\phi = [u,\,v]$.  %If we impose that $[u,\,v] \to [0,\,0]$ as $\vert x \vert \to \infty$, then we have $\tilde\phi \to c$ as $\vert x \vert \to \infty$ within the fluid bulk where $c$ is an arbitrary constant. By introducing ${\phi} = \tilde\phi - c$, we can reformulate the equations in terms of $\phi$ where $\phi\to 0$ as $\vert x \vert \to\infty$ thus eliminating any background current from our formulation.
        
    \subsection{The Nonlocal Formulation of \cite{ablowitz2006new}}
        As mentioned in the introduction, t are several different reformulation of \eqref{eqn:laplace1d}-\eqref{eqn:dynamic1d} that yield the equations of motion in terms of free-surface, and the trace of the velocity potential along the free-surface.  The recent reformulation given by \cite{ablowitz2006new} yields a coupled system of differential and integro-differential equations in terms of the physical variables $\eta(x,t)$ and $q(x,t) = \phi(x,\eta(x,t),t)$.  
        
        In their work, the authors begin by considering the velocity potential $\phi(x,z,t)$ that satisfies \eqref{eqn:laplace1d}-\eqref{eqn:dynamic1d}, and a harmonic test function $\varphi(x,z)$.  Via Green's second identity, they write the equation 
        \begin{equation}
            \cint{\varphi_z\left(\grad\phi\dotn\right) - \varphi_x\left(\grad\phi\dott\right)} = 0, \label{eqn:GreenPhiAFM}
        \end{equation}
        where $\vn$ is the outward pointing normal, and $\vt$ is the tangent vector.  We also assume that there is sufficiently fast decay of both $\phi(x,z,t)$ and $\eta(x,t)$ such that \eqref{eqn:GreenPhiAFM} makes sense.    

        Choosing the specific test functions $\varphi = e^{-ikx}\sinh(k(z+h))$ eliminates any contribution from the bottom boundary and yields the system  
        \begin{eqnarray}
            &\displaystyle \wint{e^{-ikx}\left(\eta_t\cosh(k(\eta+h)) + iq_x\sinh(k(\eta+h))\right)} = 0,\label{eqn:AFMNonlocal}\\
            &\displaystyle q_t + \frac{1}{2}q_x^2 + g\eta - \frac{1}{2}\frac{(\eta_t + q_x\eta_x)^2}{1 + \eta_x^2} = \frac{\sigma\eta_{xx}}{\rho(1 + \eta_x^2)^{3/2}}.\label{eqn:AFMLocal}
        \end{eqnarray}
        One can think of \eqref{eqn:AFMNonlocal} as an implicit relationship defining the Dirichlet-to-Neumann map at the free surface.  Indeed, \cite{haut_ablowitz_2009} show that \eqref{eqn:AFMNonlocal} can be used to generate the same expansion of the DNO as found in \cite{craig1993numerical}. They also establish the equivalence of the two formulations with that of Zakharov \cite{zakharov1968stability} and Craig and Sulem \cite{craig1993numerical}; see \cite{haut_ablowitz_2009} for details.
    %----------------------------------------------------------------------
    \subsection{Derivation of the Nonlocal/Nonlocal Formulation}\label{sec:derivation1d}
        Here, we aim to recast \eqref{eqn:AFMNonlocal}-\eqref{eqn:AFMLocal} into a general system of coupled integro-differential equations by deriving two separate nonlocal equations.  Following the work of \cite{ablowitz2006new}, we begin by introducing a harmonic test function $\varphi(x,z)$ so that via Green's second identity, we have 
        \begin{equation}
            \cint{\varphi_z\left(\grad\phi\dotn\right) - \phi\left(\grad\varphi_z\dotn\right)} = 0, \label{eqn:GreenPhi}
        \end{equation}
        where $\vn$ is the outward pointing normal.  While \eqref{eqn:GreenPhiAFM} and \eqref{eqn:GreenPhi} are equivalent up to an integration-by-parts, there are some specific advantages of working with \eqref{eqn:GreenPhi} that will be exploited in \Cref{sec:maps}.

        Imposing the kinematic boundary condition given by \Cref{eqn:kinematic1d}, \Cref{eqn:GreenPhi} becomes 
        \begin{equation}
            \sint{\varphi_z\eta_t -q(\varphi_{zz} - \eta_x\varphi_{xz})} = \bint{-Q\varphi_{zz}},\label{eqn:nonlocalA}
        \end{equation}
        in terms of $\eta(x,t)$, $q(x,t)$, and the potential at the bottom $Q(x,t) = \phi(x,-h,t)$.  Here, we have introduced the notation $$\int_{\mathscr{S}}f(x,z,t)\,dx = \int_\mathbb{R} f(x,\eta,t)\,dx \quad \text{and}\quad \int_{\mathscr{B}}f(x,z,t)\,dx = \int_{\mathbb{R}}f(x,-h,t)\,dx$$ to denote the integral over the whole line where $z = \eta(x,t)$ and $z = -h$ respectively. \Cref{eqn:nonlocalA} is the first integro-differential equation in our system of equations an will be referred to as the first nonlocal equation. It is worth noting that \Cref{eqn:nonlocalA} is a trivial generalization the nonlocal equation given in \cite{ablowitz2006new} and shown above in \eqref{eqn:AFMNonlocal}.
       
        To close our nonlocal/nonlocal formulation, we need to incorporate the dynamic boundary condition given by \Cref{eqn:dynamic1d}.  In contrast to the formulation \eqref{eqn:AFMNonlocal}-\eqref{eqn:AFMLocal}, we seek to replace \eqref{eqn:AFMLocal} with a nonlocal equation using the same test-function $\varphi$.  To achieve this, we simply take a time derivative of \eqref{eqn:nonlocalA} to find 
        \begin{equation}
            \sint{\dt\left(\varphi_z\eta_t\right) - q_t\varphi_{zz} - (q_x\eta_t - q_t\eta_x)\varphi_{xz}} = -\bint{Q_t\varphi_{zz}}.\label{eqn:nonlocalBa}
        \end{equation}
        At this point, we have yet to incorporate the dynamic boundary condition and so \eqref{eqn:nonlocalBa} is not independent from \eqref{eqn:nonlocalA}.  We can achieve this independence by  rewriting \eqref{eqn:nonlocalBa} as
        \begin{equation}
            \sint{\dt\left(\varphi_z\eta_t\right) - q_t\varphi_{zz} + (q_x\eta_t - q_t\eta_x)\varphi_{xz}-2(q_x\eta_t - q_t\eta_x)\varphi_{xz}} = -\bint{Q_t\varphi_{zz}}.
        \end{equation}
        Then, using similar ideas to those presented in \cite{olver} (see \Cref{app:simplification1d} for specifics), along with the appropriate integral theorems noting $\eta_t = \grad\phi\dotn$, we arrive at the final relationship:
        \begin{equation}
            \sint{ q_x\eta_t\varphi_{xz}-\left(\varphi_{zz}+\eta_x\varphi_{xz}\right)\left(q_t + g\eta - \dx\frac{\sigma\eta_x}{\sqrt{1+\eta_x^2}}\right) } = \frac{1}{2}\bint{Q_x^2\,\varphi_{zz}}.\label{eqn:nonlocalB}
        \end{equation}
        Together, \eqref{eqn:nonlocalA} and \eqref{eqn:nonlocalB} form a system of equations relating the free-surface variables $\eta(x,t)$ and $q(x,t)$ to the Dirichlet and Tangential Derivative values of $\phi$ evaluated at the bottom of the fluid. 
        \begin{remark}
            Alternatively, we could derive \eqref{eqn:nonlocalB} by considering Green's second identity for $\phi_t$ and $\varphi$ such that 
            \begin{equation}
                \cint{\varphi_z\left(\nabla\phi_t\dotn\right) - \phi_t\left(\nabla\varphi_z\dotn\right)} = 0.\label{eqn:GreenPhiT}
            \end{equation}
            Substituting \eqref{eqn:kinematicBottom1d}, \eqref{eqn:kinematic1d}, and \eqref{eqn:dynamic1d} into \eqref{eqn:GreenPhiT} would ultimately yield the expression.
        \end{remark}
        
        In order to consider the system closed for the unknowns $\eta(x,t)$, $q(x,t)$ and $Q(x,t)$, we need to consider a complete family of harmonic test functions $\varphi$.  One such family is the set of functions of the form $$\varphi \in\left\lbrace e^{-ikx}\cosh(k(z+h)), e^{-ikx}\sinh(k(z+h)),\, k\in\mathbb{R}\right\rbrace.$$   However, a subset of these functions yields enough information to fully close the system for the surface variables $\eta(x,t)$ and $q(x,t)$.  Choosing the harmonic function $\varphi = e^{-ikx}\sinh(k(\eta+h))$, \eqref{eqn:nonlocalA} and \eqref{eqn:nonlocalB} yield the following system of equations in the surface variable $\eta(x,t)$ and $q(x,t)$:
        \begin{eqnarray}
            &&\displaystyle \fint{\eta_t\cosh(k(\eta+h)) + iq_x\sinh(k(\eta+h)} = 0,\label{eqn:AFMNonlocalA}
        \end{eqnarray}
        and
        \begin{eqnarray}
            &&\displaystyle \fint{(q_x\eta_t - \eta_x(q_t +g\eta - \dx\frac{\sigma\eta_x}{\sqrt{1+\eta_x^2}}))\cosh(k(\eta+h))}\nonumber\\
            &&\qquad \qquad\displaystyle - \fint{i(q_t + g\eta- \dx\frac{\sigma\eta_x}{\sqrt{1+\eta_x^2}})\sinh(k(\eta+h))} = 0,\label{eqn:AFMNonlocalB}
        \end{eqnarray}
        which are valid for all $k\in\mathbb{R}$.  
        \begin{remark}
            In simplifying the above equations, we divided by $k$.  This might appear to invalidate the equations when $k = 0$.  However, in the limit as $k\to 0$, both \eqref{eqn:AFMNonlocalA} and \eqref{eqn:AFMNonlocalB} yield the known conserved densities $T_3 = \eta$ and $T_1 = -q\eta_x$ respectively as denoted in \cite{olver}. Thus, we can formulate the equations for all $k\in\mathbb{R}$.  There are deeper connections to conservation laws that will be discussed in \cite{ConLaw_Paper_1}.

            Furthermore, while it appears that \eqref{eqn:AFMNonlocalA} and \eqref{eqn:AFMNonlocalB} are equivalent to the Hamiltonian formulation given by Zakharov \cite{zakharov1968stability} and Craig and Sulem \cite{craig1993numerical}.  This has yet to be established.  Perhaps a more logical route to establish the equivalence would be via the Lagrangian formulation and variational principles due to \cite{miles1977hamilton,luke1967variational}.  This will be explored in future work. 
        \end{remark}

    % \subsection{Connection with the Lagrangian}
    %     If we consider the variational principle for water waves (see \cite{luke1967variational,miles1977hamilton}), we can derive the above non-local/non-local formulation via a variational principle.  Beginning with the Lagrangian as given in \cite{miles1977hamilton}, we have
    %     \begin{equation}
    %         \mathcal{L} = \int_{0}^T\left[\sint{\phi\eta_t - \frac{1}{2}(\eta^2 - h^2)} - \iint_{\mathscr{D}(t)}   \frac{1}{2}\vert\grad\phi\vert^2\:dz\:dx\right]\:dt.
    %     \end{equation}
    %     Taking the variation with respect to $\eta$ yields the following 
\section{Traveling Wave Solutions}

    To find the equations for traveling waves, we seek solutions that are stationary in a coordinate system moving with a constant speed $c$.  Thus, we introduce the transformation $\partial_t \to -c\partial_x$. Using that both $\eta$ and $q$ decay rapidly as $\vert x \vert \to \infty$ as well as integration by parts, \eqref{eqn:AFMNonlocalA} and \eqref{eqn:AFMNonlocalB} become
    \begin{eqnarray}
        &&\displaystyle \fint{(q_x-c)\sinh(k(\eta+h))} = 0,\label{eqn:trans1}\\
        &&\displaystyle \fint{-ck^2(q_x-c)\sinh(k(\eta+h))- (c^2 - 2g\eta)k^2\sinh(k(\eta+h))}\nonumber\\
        &&\displaystyle \qquad \qquad -\fint{gk\cosh(k(\eta+h))} = 0,~~~~\label{eqn:trans2}
    \end{eqnarray}
    where we have set the coefficient of surface tension to be zero for the remainder of the paper.  

    Substituting the relationship given by \eqref{eqn:trans1} into \eqref{eqn:trans2}, we can eliminate the velocity potential at the free surface in order to generate a single scalar equations for traveling waves in terms of $\eta$ given by 
    \begin{equation}
        \fint{(c^2 - 2g\eta)k^2\sinh(k(\eta+h)) + gk\cosh(k(\eta+h))} = 0.\label{eqn:singleNew}
    \end{equation} 
    This can be contrasted with the single equation for traveling waves given in \cite{deconinck_oliveras_2011}.  Here, we effectively substituted the kinematic boundary condition \Cref{eqn:kinematic1d} into the dynamic boundary condition \Cref{eqn:dynamic1d}.  In \cite{deconinck_oliveras_2011}, the dynamic boundary becomes a quadratic in $q_x$ for traveling wave solutions.  This is solved for $q_x$ terms of $\eta$ and substituted into \eqref{eqn:trans1} yielding
    \begin{equation}
        \fint{\sqrt{(c^2-2g\eta)(1 + \eta_x^2)}\sinh(k(\eta+h))} = 0.\label{eqn:singleOld}
    \end{equation}

   Both \eqref{eqn:singleNew} and \eqref{eqn:singleOld} represent single equations for the unknown free-surface $\eta$.  However, a curious difference between the two equations is the lack of $\eta_x$ explicitly appearing in the integrand of \eqref{eqn:singleNew}.  Similarly, both equations can be posed for periodic boundary conditions.  If we restrict to examining $2\pi$ periodic traveling-wave solutions, we can derive the analogues of \eqref{eqn:singleNew} and \eqref{eqn:singleOld} as  
    \begin{equation}
        \int_{0}^{2\pi}{e^{-ikx}\left[(c^2 - 2g\eta)k^2\sinh(k(\eta+h)) + gk\cosh(k(\eta+h))\right]\,dx} = 0, \qquad k\in \mathbb{Z},\label{eqn:singleNewPeriodic}
    \end{equation} 
    and 
    \begin{equation}
        \int_0^{2\pi}{e^{-ikx}\left[\sqrt{(c^2-2g\eta)(1 + \eta_x^2)}\sinh(k(\eta+h))\right]\,dx} = 0,\label{eqn:singleOldPeriodic}
    \end{equation}

where $k$ is restricted to the integers in order to cancel contributions from the bulk velocities at $x = 0$ and $x = 2\pi$ for both \eqref{eqn:singleNewPeriodic} and \eqref{eqn:singleOldPeriodic}.  It is worth noting that deriving the Stokes expansion for periodic traveling wave solutions using \eqref{eqn:singleNewPeriodic} is somewhat easier to compute by hand when compared to \eqref{eqn:singleOldPeriodic} due to the absence of both the square root and derivatives of the free-surface $\eta(x)$. Likewise, both equations are straight-forward to solve numerically. However, the lack of derivative in \eqref{eqn:singleNewPeriodic} may yield less numerical artifacts when computing solutions spectrally in comparison to \eqref{eqn:singleOldPeriodic}.  This warrants further exploration.  
\section{Maps between the free-surface variables and the pressure at the bottom}\label{sec:maps}
    In a similar spirit to the previous work of \cite{oliveras2012recovering,vasan2014pressure,bonneton2017recovering}, various nonlocal maps between boundary information is useful for investigating inverse problem.  One such inverse problem is reconstructing the wave profile $\eta$ from measurements of the pressure taken at the bottom of the fluid. 
    
    Returning the the whole-line problem, we can find a define a direct maps from the pressure at the bottom of the fluid domain to the free-surface variables.  For example, simply subtracting \eqref{eqn:nonlocalB} from \eqref{eqn:nonlocalBa} yields a direct map from surface variables to the pressure at the bottom of the fluid given by 
    \begin{equation}
        \bint{p_d\,\varphi_{zz}} = \rho\sint{\eta_{tt}\varphi_z+ \left(\eta_t^2 + g\eta\right)\varphi_{zz} + \left(g\eta\eta_x - 2\left(q_x\eta_t - q_t\eta_x\right)\right)\varphi_{xz}}\label{eqn:pressureMap}
    \end{equation}
    where $p_d(x,z,t) = p(x,z,t) + \rho g z$ is the dynamic pressure.  Using a suitable basis of harmonic functions $\varphi$ should allow one to reconstruct pressure at the bottom from measurements of $\eta,$ $\eta_x$, $\eta_t$, $\eta_{tt}$, $q_x$, and $q_t$.  

    In many applications such as those discussed in \cite{oliveras2012recovering}, one is often interested in the inverse problem.  As stated above, this would require significant measurements that are impracticle in application.  Given measurements of $\eta$ and $\eta_t$, one could reconstruct $q_x$ using the Normal-to-Tangential maps \cite{hOperator}.  Finally, $q_t$ can be reconstructed via \eqref{eqn:AFMLocal} and $\eta_{tt}$ via \eqref{eqn:nonlocalBa} with the specific choice that $\varphi = e^{-ikx}\sinh(k(z+h))$.  Of course, this all presumes that $\eta_x$ can also be measured and that spatial integrals can be calculated from a single time-series of the data measured at a point.  As discussed in \cite{bonneton2017recovering}, this is a challenge.  However, the techniques outlined in their paper can be used to estimate $\eta_x$ as well as the spatial components needed for integration - thus closing the loop.  
    
    While this topic will be explored in more depth in future work, we are able to easily derive various asymptotic relationships between the free-surface variables and the pressure at the bottom.  To explore this, we consider the nondimensional version\footnote{The details of the non-dimensionalization are discussed in \Cref{app:nondimensionalization}.} of \eqref{eqn:pressureMap} given by 
    \begin{equation}
        \bint{  p_d\,\varphi_{zz}} = \sint{\eta\varphi_{zz} + \mu^2\eta_{tt}\varphi_z + \epsilon\mu^2\left(\eta_t^2\varphi_{zz} + \left(\eta\eta_x + 2\left(q_t\eta_x - q_x\eta_t\right)\right)\varphi_{xz}\right)},\label{eqn:nondimPressureMap}
    \end{equation}
    where $\varphi$ now satisfies $\mu^2\varphi_{xx}+\varphi_{zz} = 0$, and the bottom and surface integrals are evaluated at $z = -1$ and $z = \epsilon\eta$ respectively.    
    
    Choosing $\varphi = e^{-ikx}\cosh(\mu k(z+1))$, we can find an expression for the spatial Fourier transform of the bottom pressure given in term of the free-surface variables.  The relationship is as follows
    \begin{eqnarray}
       &&\displaystyle\fint{\left(\eta+\epsilon\mu^2\eta_t^2\right)\mathcal{C} + \left(\frac{\mu}{k}\eta_{tt}-i\epsilon\mu\left(\eta\eta_x + 2\left(q_t\eta_x-q_x\eta_t\right)\right)\right)\mathcal{S}}\nonumber\\
       &&\displaystyle\qquad=\fint{p_{d,b}}, \label{eqn:option1Full}
    \end{eqnarray}
    where $p_{d,b}$ represents the nondimensionalized dynamic pressure evaluated at the bottom of the fluid domain, $\mathcal{S} = \sinh(\mu k (\epsilon\eta+1))$, and $\mathcal{C} = \cosh(\mu k (\epsilon\eta+1))$.
    Taking the balance $\epsilon\sim\mu^2$, \eqref{eqn:option1Full} yields the relationship between the spatial Fourier transform of $p_{d,b}$ and $\eta$:
    \begin{equation}
        \hat{p}_{d,b} =\hat{\eta}+ \epsilon\hat{\eta}_{tt} + \frac{\epsilon k^2}{2}\hat{\eta} + \mathcal{O}(\epsilon^2).
    \end{equation}
    It is worth noting that at leading order, we recover precisely the hydrostatic approximation.  
    
    While we chose to work with $\varphi = e^{-ikx}\cosh(\mu k(z+1))$, there is no reason to restrict ourselves to this particular choice.  In \Cref{tab:asyRel} we compare various asymptotic formulae generated by three slightly different choices of $\varphi$ with the same asymptotic balance $\epsilon\sim\mu^2$.  

    \begin{table}[htp]
        \caption{Comparison of various asymptotic formulae with $\epsilon \sim\mu^2$. Details are provided in \Cref{app:nondimensionalization}.}\label{tab:asyRel}
        \small{\begin{tabular}{p{.25\textwidth}p{.05\textwidth}p{.54\textwidth}r>{\raggedleft}p{.1\textwidth}}\toprule
            Choice of $\varphi$ && Resulting Model &\\
            \hline
            $\displaystyle \varphi = e^{-ikx}\cosh(\mu k (z + 1))$ && $\displaystyle p_{d,b} = \eta + \epsilon\left(\eta_{tt} -\frac{1}{2}\eta_{xx}\right) + \mathcal{O}(\epsilon^2)$&\refstepcounter{equation}\label{eqn:option1asy}(\arabic{equation})\\
            &&&\\
            $\displaystyle \varphi = e^{-ikx}\cosh(\mu kz)$ && $\displaystyle \left(1 - \frac{\epsilon}{2}\partial_x^2\right)p_{d,b} = \eta + \mathcal{O}(\epsilon^2)$&\refstepcounter{equation}\label{eqn:option2asy}(\arabic{equation})\\ && \\
            $\displaystyle \varphi = e^{-ikx}\sinh(\mu kz)$ && $\displaystyle \left(1 - \frac{\epsilon}{6}\partial_x^2\right)p_{d,b} = \eta - \left(\frac{1}{2}\eta^2 + \epsilon\left(\partial_x^{-1}\eta_t\right)^2\right) + \mathcal{O}(\epsilon^2)$&\refstepcounter{equation}\label{eqn:option3asy}(\arabic{equation})\\\bottomrule
        \end{tabular}}
    \end{table}
    
    While it may seem as though \eqref{eqn:option1asy}, \eqref{eqn:option2asy}, and \eqref{eqn:option3asy} represent a system of overdetermined equations for $\eta$ and $p_{d,b}$, a careful inspection shows that they are consistent provided $$\eta_{tt} - \eta_{xx} = \epsilon\partial_x^2\left[\frac{1}{3}\eta_{xx} + \frac{1}{2}\eta^2 + \left(\partial_x^{-1}\eta_t\right)^2\right] + \mathcal{O}(\epsilon^2).$$ This is precisely the Boussinesq equation for the free-surface $\eta$ which is valid to the same order (see \cite{ablowitz2011nonlinear} for details).   

    While (\ref{eqn:option1asy}-\ref{eqn:option3asy}) are all valid to the same order, they contain different challenges.  For example, \eqref{eqn:option1asy} provides a direct linear map from $\eta$ to $p_{d,b}$, while \eqref{eqn:option2asy} provides a direct linear map from $p_{d,b}$ to $\eta$.  Both formulae are valid at the same asymptotic order, however, \eqref{eqn:option3asy} contains both nonlinear and nonlocal terms in $\eta$ and may be challenging to use when working with real data.

\section{Concluding Remarks}
In summary, this short paper outlines a new formulation of the water-wave problem for irrotational waves with a one-dimensional surface. Connections to other ideas such as conservation laws, Lagrangian formulations, inverse problems, and extensions to other fluid configurations (two-dimensional free-surfaces, density stratification, and vorticity) will be explored in forthcoming papers. 

\subsection*{Acknowledgements}
    The author would like to Vishal Vasan for fruitful discussions related to this work.  Likewise, the author would like to acknowledge the assistance of Salvatore Calatola-Young, Gabriel Greenstein, and Reece Keller for double checking the derivations and pointing out sign errors.

    This material is based upon work supported by the National Science Foundation under Grant Number DMS-1715082. Any opinions, findings, and conclusions or recommendations expressed in this material are those of the author and do not necessarily reflect the views of the National Science Foundation.
%%%%%%%%%%%%%%%%%%%%%%%%%%%%%%%%%%%%%%%%%%%%%%%%%%%%%%%%%%%%%%%%%%%%%%%%

%%%%%%%%%%%%%%%%%%%%%%%%%%%%%%%%%%%%%%%%%%%%%%%%%%%%%%%%%%%%%%%%%%%%%%%%%%
\appendix
\section{Simplification Formulae for Irrotational Fluids}\label{app:simplification1d}
    Consider harmonic functions $\varphi(x,z)$, $A(x,z)$, and $B(x,z)$ such that $A$ and $B$ decay sufficiently as $\vert x \vert \to \infty$ such that the following integrals make sense.  Then, via the divergence theorem on our domain $\mathscr{D}$, we find
     
    \begin{equation}
        \sint{A\normD{\varphi_z} + B\normD{\varphi_x}} = \bint{A\varphi_{zz} + B\varphi_{xz}}. \label{eqn:simClosedForm}
    \end{equation} 

   Using both the kinematic and dynamic boundary conditions given by \eqref{eqn:kinematic1d} and \eqref{eqn:dynamic1d} respectively, the following relationship is straightforward to derive: 
    \begin{eqnarray*}
        &&\sint{-q_t \varphi_{zz} + (\eta_tq_x - \eta_xq_t)\varphi_{xz}}\\%&=& \sint{-(\phi_t + \eta_t\phi_z)\varphi_{zz} + (\eta_t(\phi_x+\eta_x\phi_z) - \eta_x(\phi_t+\eta_x\phi_z))\varphi_{xz}}%\\
        % &=&\sint{(-\phi_t - \phi_z^2 + \eta_x\phi_x\phi_z)\varphi_{zz} + (\phi_x\phi_z - \eta_x\phi_x^2 - \phi_t\eta_x)\varphi_{xz}}\\
        % &=&\sint{(\frac{1}{2}(\phi_x^2 - \phi_z^2) + g\eta + \eta_x\phi_x\phi_z)\varphi_{zz} + (\phi_x\phi_z - \eta_x\frac{1}{2}(\phi_x^2-\phi_z^2 - g\eta))\varphi_{xz}}\\
        &&\qquad\qquad=\sint{\left(g\eta -\dx\frac{\sigma\eta_x}{\sqrt{1+\eta_x^2}}\right)(\varphi_{zz} + \eta_x\varphi_{xz}) + A\normD{\varphi_z} + B\normD{\varphi_x}},
    \end{eqnarray*}
    where $A$ and $B$ are given by $$A = \frac{1}{2}\left(\phi_x^2 - \phi_z^2\right), \qquad B = \phi_x\phi_z.$$
    It is straight-forward to show that both $A$ and $B$ are harmonic.  Thus, under suitable boundary conditions for $\eta(x,t)$ and $\phi(x,z,t)$ in $x$ (sufficient decay on the whole line, or periodicity in $x$), we find
    \begin{eqnarray*}
        &&\sint{-q_t \varphi_{zz} + (\eta_tq_x - \eta_xq_t)\varphi_{xz}}\\
        &&\qquad \qquad=\sint{\left(g\eta -\dx\frac{\sigma\eta_x}{\sqrt{1+\eta_x^2}}\right)(\varphi_{zz} + \eta_x\varphi_{xz}) }%
        + \frac{1}{2}\bint{Q_x^2\varphi_{zz}}
    \end{eqnarray*}
    where we have used \eqref{eqn:kinematicBottom1d} along with the notation $Q(x,t) = \phi(x,-h,t)$ to simplify the integrand evaluated at the bottom.

\section{Dimensionless Variables}\label{app:nondimensionalization}
    We introduce the nondimensional quantities as follows:
    \begin{eqnarray*}
        &\displaystyle x = x^*L, \quad z = z^* h, \quad t = \frac{L}{c_0}t^*, \quad c_0 = \sqrt{gh},\quad \epsilon = \frac{a}{h}, \quad \mu = \frac{h}{L},\\&\\
        &\displaystyle \eta = a\eta^*,\qquad \frac{\rho gz + p}{\rho} = \epsilon p_d^*,\qquad Q = \frac{Lga}{c_0}Q^*,\qquad q = \frac{Lga}{c_0}q^*,
    \end{eqnarray*}
    where $L$ is a typical horizontal length scale, $h$ is the undisturbed fluid depth, $g$ is the gravitational constant, and $a$ is the amplitude of the surface wave. With this change of variables and omitting the $^*$'s, the nondimensional version of \eqref{eqn:pressureMap} is  
    \begin{equation}
        \bint{  p_d\,\varphi_{zz}} = \sint{\eta\varphi_{zz} + \mu^2\eta_{tt}\varphi_z + \epsilon\mu^2\left(\eta_t^2\varphi_{zz} + \left(\eta\eta_x + 2\left(q_t\eta_x - q_x\eta_t\right)\right)\varphi_{xz}\right)},
    \end{equation}    
    where $\varphi(x,z)$ is now a function satisfying $\mu^2\varphi_{xx} + \varphi_{zz} = 0$, the bottom integral is evaluated such that $z = -1$, and the surface integral is evaluated where $z = \epsilon\eta$.

\bibliographystyle{alpha}
\bibliography{./common/references}
\end{document}